\documentclass[
    ,final            
  ]
  {aipproc}

\layoutstyle{6x9}
\usepackage{epsf,epsfig,psfrag}
\usepackage{amsmath}

\def\CR{\nonumber\\}

\def\slash#1{#1 \hskip-0.5em /}

\def\calO{\mathcal{O}}
\def\as{\alpha_s}
\def\beq{\begin{align}}
\def\eeq{\end{align}}
\def\vareps{\varepsilon}
\def\no{\nonumber}

\def\om{\omega}

\def\as{\alpha_s}
\def\calO{\mathcal{O}}

\def\be{\begin{equation}}
\def\ee{\end{equation}}

\def\om{\omega}

\begin{document}

\title{Light-Cone Distribution Amplitudes\\[0.5em]
for Non-Relativistic Bound States
}

{\small \hfill \tt Preprint SI-HEP-2007-08, TTP-07-35}

\author{Th.~Feldmann}{
  address={Fachbereich Physik,
Theoretische Physik I,\\
Universit\"at Siegen,
Emmy Noether Campus, D-57068 Siegen, Germany}
}

\author{G.~Bell}{
  address={Institut f\"ur Theoretische Teilchenphysik,\\
Universit\"at Karlsruhe, 
D-76128 Karlsruhe, Germany}
}

\begin{abstract}
We calculate light-cone distribution amplitudes for
non-relativistic bound states, including radiative
corrections from relativistic gluon exchange to
first order in the strong coupling constant. 
Our results apply to hard exclusive reactions
with non-relativistic bound states in the QCD factorization
approach like, for instance, 
$B_c \to \eta_c \ell\nu$ or $e^+e^- \to J/\psi \eta_c$. 
They also serve as a toy model for light-cone distribution
amplitudes of light mesons or heavy $B$ and $D$ mesons.
\end{abstract}

\maketitle



\section{Introduction}
\label{sec:intro}

Exclusive hadron reactions with large momentum transfer involve
strong interaction dynamics at very different momentum scales. 
In cases where the hard-scattering process is dominated by light-like
distances, the long-distance hadronic information is given in terms of
so-called light-cone distribution amplitudes (LCDAs) which are
defined from hadron-to-vacuum matrix elements of non-local operators
with quark and gluon field operators separated along the light-cone.
Representing universal hadronic properties,
LCDAs can either be extracted from experimental data, or they have to
be constrained by non-perturbative methods. The most extensively
studied and probably best understood case is the leading-twist
pion LCDA, for which experimental constraints from the $\pi-\gamma$
transition form factor \cite{Gronberg:1997fj}, as well as estimates for
the lowest moments from QCD sum rules \cite{Ball:1998je,Khodjamirian:2004ga,Ball:2006wn} and lattice QCD
\cite{lattice}
exist. On the other hand, our knowledge on LCDAs for heavy $B$ mesons
\cite{Grozin:1996pq,Lange:2003ff,Braun:2003wx},
and even more so for heavy quarkonia \cite{Ma:2006hc,Braguta:2006wr}, 
had been relatively poor until recently.

The situation becomes somewhat simpler, if the hadron under
consideration can be approximated as a non-relativistic bound
state of two sufficiently heavy quarks. In this case we expect
exclusive matrix elements -- like transition form factors
\cite{Bell:2006tz} and, in particular, the LCDAs -- 
to be calculable perturbatively, since the quark masses
provide an intrinsic physical infrared regulator. 
In these proceedings we report about results from 
\cite{Bell:2007}, where we have calculated the LCDAs for 
non-relativistic meson bound 
states including relativistic QCD corrections to first order
in the strong coupling constant at the non-relativistic
matching scale which is set by the mass of the
lighter quark in the hadron. 
%


\section{Light-cone distribution amplitudes}

\label{sec:NRapprox}

The wave function for a non-relativistic (NR) bound state of a quark
and an antiquark can be obtained from the solution of the
Schr\"odinger equation with the QCD Coulomb potential.
To first approximation it describes
a quark with momentum $m_1 \, v_\mu$ and an antiquark with momentum $m_2
v_\mu$, where $v_\mu$ is the four-velocity of the meson. 
The spinor degrees of freedom for a non-relativistic
pseudoscalar bound state
are represented by the Dirac projector
$\frac12 (1+\slash{v})\gamma_5$. 
The non-relativistic approximation can also serve as a toy model
for bound states of light (relativistic) quarks. We will in the
following refer to ``heavy mesons'' as "$B$" (where we mean the realistic
example of a $B_c$ meson, or the toy model for a $B_q$ meson,
with $m_1 \gg m_2$)
and ``light mesons'' as "$\pi$" (where the realistic example is $\eta_c$, and the toy-model application would be the pion, with
$m_1\approx m_2$).


\subsection{Definition of LCDAs for light pseudoscalar mesons}

Following~\cite{Braun:1989iv,Ball:1998je} we define the 2-particle LCDAs
of a light pseudoscalar meson via  
\begin{align}
\langle \pi(P) | \bar{q}_1(y) \,  \gamma_\mu \gamma_5 \, q_2(x) | 0\rangle
&{}= -i f_\pi  \int_0^1 \! du \;
e^{i (u \, p \cdot y+\bar{u} \, p \cdot x)} \; 
\left[ p_\mu \, \phi_\pi(u) + \frac{m_\pi^2}{2 \, p\cdot z}\,z_\mu\, g_\pi(u)\right],
\CR
\langle \pi(P) | \bar{q}_1(y) \,  i \gamma_5 \, q_2(x) | 0 \rangle
&{}= f_\pi \, \mu_\pi  \int_0^1 \! du \;
e^{i (u \, p \cdot y+\bar{u} \, p \cdot x)} \; \phi_p(u) , 
\CR
\langle \pi(P) | \bar{q}_1(y) \,  \sigma_{\mu \nu} \gamma_5 \, q_2(x) |0\rangle
&{}= i f_\pi\, \tilde{\mu}_\pi
(p_\mu z_\nu - p_\nu z_\mu)
\int_0^1 \! du \;
e^{i (u \, p \cdot y+\bar{u} \, p \cdot x)} \;
\frac{\phi_\sigma(u)}{2D-2}
\label{eq:Pion:def} 
\end{align}
with two light-like vectors $z_\mu=y_\mu-x_\mu$ and $p_\mu=P_\mu-m_\pi^2/(2 P \, \cdot z)\,z_\mu$, with the usual gauge link factor $[y,x]$ (Wilson line) understood implicitly.
Here $u = 1-\bar u$ denotes the light-cone momentum fraction of the quark
in the pion,
with $\phi_\pi(u)$ being the twist-2 LCDA, 
while $\phi_p(u)$ and $\phi_\sigma(u)$ are of twist-3.
For completeness, we have also quoted the twist-4 LCDA
$g_\pi(u)$ which, like the 3-particle LCDAs, will not be considered further.
All LCDAs are normalized to 1, such that the prefactors
in (\ref{eq:Pion:def}) are defined in the local limit $x \to y$.
In the definition of $\phi_\sigma(u)$, we have included a factor $3/(D-1)$, such
that the relation between $\mu_\pi$ and $\tilde \mu_\pi$ from the equations of motion (eom),
\begin{align}
\tilde \mu_\pi = \mu_\pi - (m_1+m_2) \,,
\label{eq:eom:local1}
\end{align}
is maintained in $D\neq 4$ dimensions.
In the local limit the eom further imply
\begin{align} 
\mu_\pi &{}= \frac{m_\pi^2}{m_1 + m_2}, 
\qquad
\int_0^1 du \, u \; \phi_p(u){}= \frac12 + \frac{m_1-m_2}{2\mu_\pi}.
\label{eq:eom:local2}
\end{align}
Notice that (\ref{eq:eom:local1},\ref{eq:eom:local2}) hold for the \emph{bare} parameters and distribution amplitudes.

At tree level, and in leading order of the expansion in the
non-relativistic velocities, the two quarks in the non-relativistic
wave function simply share the momentum of the meson according to their
masses, $p_i^\mu \simeq m_i/(m_1+m_2) \, P^\mu$.
For ``light mesons'' this implies
\begin{align}
  \phi_\pi(u) &{} \simeq \phi_p(u)  \simeq g_\pi(u) \simeq \delta(u- u_0) \,,
\label{delta}
\end{align}
where $u_0=m_1/(m_1+m_2)$. Consequently, all positive and negative moments
of the distribution amplitudes
are simply given in terms of the corresponding power of $u_0$. 
Notice that $\tilde \mu_\pi \simeq 0$ at tree-level, and the corresponding
LCDA $\phi_\sigma(u)$ can only be determined by considering the corresponding
one-loop expressions (see \cite{Bell:2007}).


\subsection{Definition of LCDAs for heavy pseudoscalar mesons}

We define the 2-particle LCDAs of a heavy pseudoscalar $B$ meson following
\cite{Grozin:1996pq,Beneke:2000wa},
\begin{align}
\langle 0 | \bar{q}^\beta(z) \,  h_v^\alpha(0) | B(v) \rangle
 & {} = - \frac{i \hat{f}_B(\mu) M}{4} \left[ \frac{1
+\slash{v}}{2} \left\{ 2 \tilde{\phi}_B^+(t) +
\frac{\tilde{\phi}^-_B(t)-\tilde{\phi}_B^+(t)}{t} \slash{z}
\right\} \gamma_5 \right]^{\alpha \beta} \,,
\label{eq:NRdefLCDABc}
\end{align}
where $v^\mu$ is the heavy meson's velocity, $t\equiv v \cdot z$ and $z^2=0$.
Here $\hat{f}_B$ is the (renormalization-scale dependent) decay constant in HQET. 
The Fourier-transformed expressions, which usually
appear in factorization formulas, are given through
\begin{align}
\tilde{\phi}^\pm_B(t) & {}= \int_0^\infty \! d\omega \; e^{-i\omega t}
\phi^\pm_B(\omega)\,,
\end{align}
where $\omega$ denotes the light-cone energy of the light quark
in the $B$ meson rest frame.


Including a finite spectator quark mass $m$ and
the effect of the 3-particle LCDAs $\Psi_A, \Psi_V$ 
as defined in~\cite{Kawamura:2001jm}, 
the eom become
\begin{align}
& \om\, \phi_B^-(\om)- m\, \phi_B^+(\om) + \frac{D-2}{2} 
\int_0^\om d\eta \; \left[\phi_B^+(\eta)-\phi_B^-(\eta)\right] \CR
& \qquad  =\;
(D-2) \int_0^\om d\eta  \; \int_{\om-\eta}^\infty \frac{d\xi}{\xi} \;
\frac{\partial}{\partial \xi} \; 
\left[ \Psi_A(\eta,\xi)-\Psi_V(\eta,\xi)\right] \,,
\label{eom1}
 \end{align}
which is trivially fulfilled at tree-level, 
where
$  \phi_B^+(\omega)   \simeq \phi_B^-(\omega) \simeq \delta(\omega - m) $,
and $\Psi_{V,A}(\eta,\xi) = {\cal O}(\alpha_s)$.
We have shown in \cite{Bell:2007} that 
this relation
 also holds after including $\alpha_s$ corrections to the NR limit.
(A second relation, which has been
presented in \cite{Kawamura:2001jm} and extended
here to the case $m\neq 0$, is found to be
not valid beyond tree-level.)
Moreover, at tree level, the moments of the ``heavy meson's'' LCDAs
can be related to matrix elements of local operators in HQET
\cite{Grozin:1996pq}.


\section{Relativistic corrections at one-loop}

\label{sec4}

\begin{figure}[t!]
\centerline{\parbox{13cm}{
\centerline{\includegraphics[width=13cm]{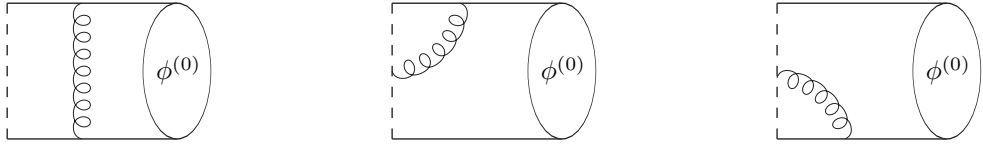}}
\caption{\label{fig:relcorr} \small \textit{Relativistic
corrections to the light-cone distribution amplitudes. The dashed
line indicates the Wilson line in the definition of the LCDAs.}}}}\vspace{2mm}
\end{figure}

The NR bound states are described by parton configurations with
fixed momenta. Relativistic gluon exchange as 
in Fig.~\ref{fig:relcorr}
leads to modifications: First, there is a correction from matching 
QCD (or, in the case of heavy mesons, the corresponding low-energy effective 
theory HQET) on the NR theory,
\begin{align}
 \phi_M &{}  = \phi_M^{(0)} + \frac{\alpha_s C_F}{4\pi} \, \phi_M^{(1)} + 
  {\cal O}(\alpha_s^2) \,.
\end{align}
Secondly, there is the usual evolution under the change
of the renormalization scale \cite{ERBL,Lange:2003ff}. In particular, the
support region for the parton momenta is extended to $0\leq u \leq 1$
for light mesons and $0 \leq \omega < \infty\,$ for heavy mesons,
respectively.

\subsection{Light mesons}

We first consider the leading-order relativistic corrections to the local
matrix elements.  We will focus
on the case of equal quark masses (results for $m_1 \neq m_2$ can be
found in \cite{Bell:2007}). Our result for the decay constant,
\begin{align}
f_\pi &{}= f_\pi^\text{NR}
\left[ 1 - 6 \, \frac{\as C_F}{4\pi}  + \calO(\as^2) \right],
\label{eq:fpires}
\end{align}
is in agreement
with~\cite{Braaten:1995ej},
and the result for 
$\mu_\pi$ and $\tilde{\mu}_\pi$ is consistent with the eom-constraint
in (\ref{eq:eom:local1},\ref{eq:eom:local2}), using
$m_\pi \simeq m_1^\text{os}+m_2^\text{os}$ in the on-shell scheme.

The remaining contributions to the NLO correction to the leading-twist 
LCDA contain an UV-divergent piece,
\begin{align}
\phi_\pi^{(1)}(u) \big|_{\rm div.} &{}= \frac{2}{\vareps} \, \int_0^1 dv \, V(u,v) \, \phi^{(0)}(v)\,,
\end{align}
which involves the well-known Efremov-Radyushkin-Brodsky-Lepage (ERBL)
evolution kernel
\cite{ERBL},
and a finite term, 
\begin{align}
\phi_\pi^{(1)}(u;\mu) &{}= 4  \left\{
\left( \ln \frac{\mu^2}{m_\pi^2 \, (1/2-u)^2} -1 \right)\!
 \left[ \left( 1+\frac{1}{1/2-u} \right) u\;\theta(1/2-u) + 
 (u \leftrightarrow \bar u) \right] \right\}_{+} \no \\
& \quad 
+4  \left\{ \frac{u (1-u)}{(1/2-u)^2} 
\right\}_{++}
 \label{eq:NRLCDABNLO} \,.
\end{align}
Here the plus-distributions are defined as
\begin{align} 
\int_0^1 \! du \; \big\{\ldots\big\}_{+} \;f(u) & {} \equiv \int_0^1 \! du \;
\big\{\ldots\big\} \; \bigg( f(u) - f(1/2) \bigg) \,,
\nonumber
\\[0.2em]
\int_0^1 \! du \; \big\{\ldots\big\}_{++} \;f(u) &{} \equiv \int_0^1 \! du \;
\big\{\ldots\big\} \; \bigg( f(u) - f(1/2) -f'(1/2) \, (u-1/2)\bigg)\,.
\label{eq:plusdef}
\end{align}

An independent calculation of the
leading-twist LCDAs for the $\eta_c$ and $J/\psi$ meson 
has been presented in~\cite{Ma:2006hc}.
Our result is not in complete agreement with their findings. In
particular, we find that the LCDA
quoted in~\cite{Ma:2006hc} is not normalized to unity as it should be.

At the non-relativistic scale, $\mu \simeq m$, the usual
expansion of $\phi_\pi(u)$ into Gegenbauer polynomials (the
eigenfunctions of the leading-order ERBL evolution equations),
does not converge very well, i.e.\ the Gegenbauer coefficients
$a_n$ drop off slower than $1/n$.
A better characterization of the LCDAs at NLO is given in terms
of the moments
\begin{align}
 \langle \xi^n \rangle_\pi \equiv &
\int_0^1 du \, (2u-1)^n \, \phi_\pi(u) \,,
\end{align}
which are linear combinations of Gegenbauer coefficients
of order $\leq n$ .
This corresponds to an expansion in terms of
$\delta$\/-function and its derivatives,
\begin{align}
 \phi_\pi(u) & {} = 2 \, \sum_n \, \langle \xi^n \rangle_\pi \,
  \frac{(-1)^n}{n!} \, \delta^{(n)}(2u-1) \,.
\end{align}
Results for the first few moments $\langle \xi^n \rangle_\pi$
are shown in Table~\ref{tab:ximom}. 
(Notice that the moments $\langle \xi^n \rangle_\pi$ also
receive corrections from sub-leading terms 
in the non-relativistic expansion,
see \cite{Braguta:2006wr}.)
 
\begin{table}[t!!bp]
\begin{tabular}{l | c c c c c}
\hline
 {$\mathbf n$} & {\bf 2}  & {\bf 4}  & {\bf 6}  & {\bf 8} & {\bf 10}
\\
\hline
 NLO ($\mu=m$, in units of $\alpha_s$) 
 & 0.333 & 0.053 & 0.019 & 0.009 & 0.005
\\
\hline
 LL ($\eta=1/5$) 
 &0.126 & 0.048 & 0.025 & 0.015 & 0.010
\\
 LL ($\eta=1/25$) 
 & 0.173 & 0.070 & 0.038 & 0.024 &0.016
\\
asymptotic 
& 0.200 & 0.086 & 0.048 & 0.030 & 0.021
\\ \hline
\end{tabular}
\caption{Convergence of
  $\langle \xi^n \rangle_\pi$ moments ($\eta=\alpha_s(\mu)/\alpha_s(m)$). \label{tab:ximom}}
\end{table}

\begin{figure}[t!!bp]
 \psfrag{phipi}{\footnotesize $\phi_\pi(u,\mu)$}
 \psfrag{u}{\footnotesize $u$}
 (a) \hspace{-1em} \parbox[c]{0.44\textwidth}{
\includegraphics[width=0.43\textwidth]{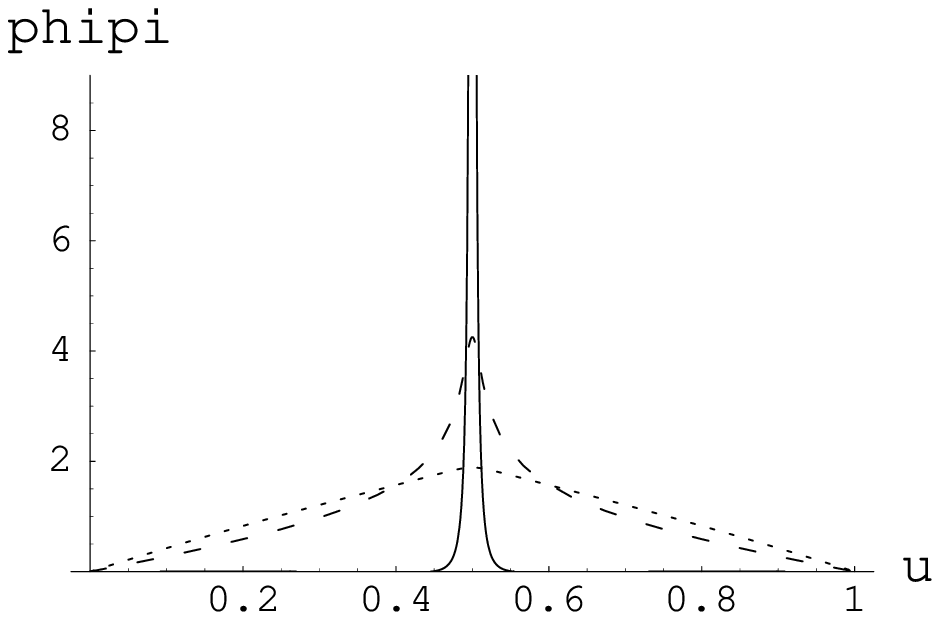}
} 
 \psfrag{phiplus}{\parbox{7em}{\footnotesize $m \, \phi_B^+(\omega,\mu)$\\ \ }}
 \psfrag{om}{\footnotesize $\omega/m$}
(b) \hspace{-.5em} 
\parbox[c]{0.44\textwidth}{ \vspace{0.3em}
\includegraphics[width=0.43\textwidth]{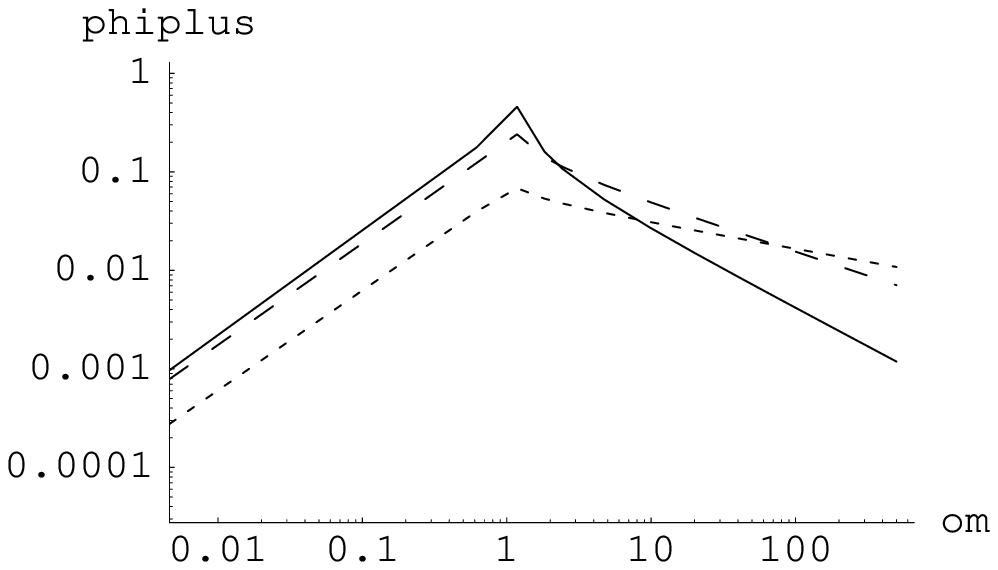} 
}
\caption{\label{fig:phiM}
(a) Solid line: approximation for $\phi_\pi(u)=\delta(u-1/2)$ in terms of 
the parametrization (\ref{model}), and its evolution
for $\eta = 1/5$ (dashed line) and
$\eta=1/25$ (dotted line).
(b) Evolution of the heavy-meson LCDA $\phi_B^+(\omega)=\delta(\omega-m)$. 
The three curves (solid, long-dashed, short-dashed)
correspond to $\eta=1/2,1/5,1/10$.}
\end{figure}

Table~\ref{tab:ximom} also
contains the moments generated by ERBL evolution of the LO
result (\ref{delta}) for two values of $\eta=\alpha_s(\mu)/\alpha_s(m)$
and in the asymptotic limit.
To illustrate the change of the shape of $\phi_\pi(u)$ under evolution,
we employ a parametrization which is obtained from a slight modification 
of the strategy developed in \cite{Ball:2005ei},
\begin{align}
  \phi_\pi(u) & {} \equiv   \frac{3 u \bar u}{\Gamma[a,-\ln t_c]}
  \, \int_0^{t_c} dt \, (- \ln t)^{a-1} 
  \left( f(2u-1,i t^{1/b})+ f(2u-1,-i t^{1/b}) \right) \,.
\label{model}
\end{align}
It involves three real parameters $a > 0$, $b>0$, $t_c \leq 1$,
which can be fitted to the first three moments $\langle \xi^{2,4,6}\rangle$,
and the generating function for the Gegenbauer polynomials,
\begin{align}
  f(\xi,\theta) &{}= \frac{1}{(1 - 2 \xi \theta + \theta^2)^{3/2}} = \sum_{n=0}^\infty
  C_n^{3/2}(\xi) \, \theta^n \,.
\end{align}
Fig.~\ref{fig:phiM}(a) shows the evolution of the model LCDA
as a function of $u$. For $\eta=1/5$ the functional form still
``remembers'' the non-relativistic profile, while for $\eta=1/25$
the LCDA gets close to the asymptotic form.

The twist-3 LCDAs for the 2-particle Fock states are obtained in
the same way as the twist-2 one. 
Details can be found in \cite{Bell:2007}.
In particular, all our results are in manifest
agreement with the eom-constraint from
(\ref{eq:eom:local2}).


\subsection{Heavy mesons}

The calculation of the LCDA for a heavy meson 
goes along the same lines as for the light-meson case. 
However, important differences
arise because the heavier quark is to be treated in HQET which
modifies the divergence structure of the loop integrals. As a
consequence, the evolution equations for the LCDA of heavy mesons
\cite{Lange:2003ff} differ from those of light mesons.

Let us focus on the distribution amplitude
$\phi_B^+(\om)$ which enters the QCD factorization formulas
for exclusive heavy-to-light decays.
In the local limit we derive the corrections from soft gluon exchange
to the decay constant in HQET,
\begin{align}
\hat{f}_M(\mu) &{} = f_M^\text{NR} \left[ 1 + \, \frac{\alpha_s C_F}{4\pi}
\left( 3\ln\frac{\mu}{m}-4 \right)+ {\cal O}(\alpha_s^2) \right].
\end{align}
Notice that the decay constant of a heavy meson exhibits the
well-known scale dependence \cite{Neubert:1993mb}.
The remaining NLO corrections to the distribution amplitude $\phi_B^+(\omega)$ 
contain an UV-divergent piece, which involves
the Lange-Neubert evolution kernel $\gamma_+(\omega,\omega',\mu)$
\cite{Lange:2003ff}.
As has been shown by Lee and Neubert \cite{Lee:2005gz}, 
the solution of the evolution equation can be written in
closed form. The resulting evolution of $\phi_B^+(\omega,\mu)$
is illustrated in Fig.~\ref{fig:phiM}(b),
starting from $\delta(\omega-m)$ at the non-relativistic scale,
for three different values of $\eta=\alpha_s(\mu)/\alpha_s(m)$ 
and taking $\alpha_s(m)=1$.
From the double-logarithmic plot we can read off the asymptotic 
behaviour of the LCDA for $\omega \to 0$ and $\omega\to\infty$.
As argued on general grounds \cite{Lange:2003ff}, $\phi_B^+(\omega)$
develops a linear behaviour for $\omega \to 0$, whereas
for $\omega \to \infty$ the evolution generates a radiative
tail which tends to fall off slower than $1/\omega$ at higher scales.

The finite NLO correction to $\phi_B^+(\omega)$ reads
\begin{align} 
&\frac{\phi^{(+,1)}_B(\om;\mu)}{\om} 
 {}= 
2  \left[ \left(\ln \left[\frac{\mu^2}{(\om-m)^2}\right]-1\right) \left(
\frac{\theta(m-\om)}{m (m-\om)} 
+ \frac{\theta(\om-m)}{\om(\om-m)}\right)
\right]_+ 
\no \\
& {}
{} + 
4  \left[ \frac{\theta(2m-\om)}{(\omega-m)^2} \right]_{++} 
+ \frac{4 \, \theta(\om-2m)}{(\om-m)^2} 
- \frac{\delta(\om-m)}{m} \left( \frac12 \ln^2
\frac{\mu^2}{m^2} -  \ln \frac{\mu^2}{m^2}  +
\frac{3\pi^2}{4} + 2 \right) 
\label{phiplus}
\end{align}
with an analogous definition of plus-distributions as
in~(\ref{eq:plusdef}).
In contrast to the light-meson case, the normalization of the heavy
meson distribution amplitude is ill-defined. Imposing a hard cutoff
$\Lambda_{\rm UV}\gg m$ and expanding to first order 
in $m/\Lambda_{\rm UV}$,
we derive
\begin{align}
\int_0^{\Lambda_{\rm UV}} d\om \; \phi_B^+(\om;\mu)
& {} \simeq 1 - \frac{\alpha_s C_F}{4\pi} \left[ \frac12 \ln^2
\frac{\mu^2}{\Lambda_{\rm UV}^2}+\ln \frac{\mu^2}{\Lambda_{\rm UV}^2} +
\frac{\pi^2}{12} \right] + \ldots 
\\[0.2em]
\int_0^{\Lambda_{\rm UV}} d\om \; \om\, \phi_B^+(\om;\mu)
& {} \simeq \frac{\alpha_s C_F}{4\pi} \left[ 2\ln \frac{\mu^2}{\Lambda_{\rm UV}^2}
+6 \right] \, \Lambda_{\rm UV}
+ \ldots
\end{align}
The last two expressions provide model-independent properties of the
distribution amplitude which agree with the general results in~\cite{Lee:2005gz}. 
The two phenomenologically relevant
moments in the factorization approach to heavy-to-light decays
read
\begin{align} 
\lambda_B^{-1}(\mu) & {} \equiv \int_0^\infty d\om \;\;
\frac{\phi_B^+(\om;\mu)}{\om}  = \frac{1}{m} \left(
1 -\frac{\alpha_s C_F}{4\pi} \left[ \frac12
\ln^2 \frac{\mu^2}{m^2} - \ln \frac{\mu^2}{m^2} +\frac{3\pi^2}{4} -2
\right] \right) \,,
\nonumber \\
\sigma_B(\mu) & {} \equiv \lambda_B(\mu) \, \int_0^\infty d\om \,
\frac{\phi_B^{+}(\om;\mu)}{\om} \, \ln \frac{\mu}{\omega} 
 = \ln \frac{\mu}{m} 
+ \frac{\alpha_s C_F}{4\pi} \left[8 \zeta(3)\right] \,.
\end{align}

A similar analysis can be performed for the LCDA $\phi_B^-(\omega)$,
for details we refer to \cite{Bell:2007}. In particular, we can read 
off the anomalous dimension, 
\begin{align}
  \gamma_-^{(1)}(\omega,\om';\mu)   = {} &
\left( 4 \, \ln \frac{\mu}{\omega} - 2
\right) \delta(\omega-\om')
- 4 \, \frac{\theta(\om'-\omega)}{\om'}
\nonumber \\[0.2em]
 & {} - 4 \, \omega \left[\frac{\theta(\om'-\omega)}{\om'(\om'-\omega)} \right]_+
- 4 \, \omega \left[
\frac{\theta(\omega-\om')}{\omega(\omega-\om')} \right]_+ \,,
\label{gammam}
\end{align}
which describes the evolution of $\phi_B^-(\omega,\mu)$ in the 
Wandzura-Wilczek approximation, where 3-particle LCDAs 
(and the light quark mass $m$) are neglected. 
Among others, $\gamma_-$ is needed
to show the factorization of correlation functions in SCET sum rules
for heavy-to-light form factors \cite{DeFazio:2005dx}.
Another new result are the first positive moments of $\phi_B^-(\omega)$ 
as a function of the UV cutoff,
\begin{align}
\int_0^{\Lambda_{\rm UV}} d\om \; \phi_B^-(\om;\mu)
&{}\simeq 1 - \frac{\alpha_s C_F}{4\pi} \left[ \frac12 \ln^2
\frac{\mu^2}{\Lambda_{\rm UV}^2} - \ln \frac{\mu^2}{\Lambda_{\rm UV}^2} +
\frac{\pi^2}{12} \right] + \ldots
\\[0.2em]
\int_0^{\Lambda_{\rm UV}} d\om \; \om\, \phi_B^-(\om;\mu)
&{}\simeq \frac{\alpha_s C_F}{4\pi} \left[ 2\ln \frac{\mu^2}{\Lambda_{\rm UV}^2}
+2 \right]\, \Lambda_{\rm UV} + \ldots
\end{align}
where the divergent pieces again are expected to be model-independent.


\section{Summary}

Non-relativistic $q\bar q$ bound states have been used as
a starting point to construct light-cone distribution amplitudes
in QCD and in HQET. We considered relativistic gluon corrections
at NLO in the strong coupling at the non-relativistic scale, as
well as the leading logarithmic evolution towards higher
scales needed in QCD factorization theorems. We also studied
certain model-independent properties of light and heavy LCDAs,
including new results for the LCDA $\phi_-^B(\omega,\mu)$
in HQET.

\begin{theacknowledgments}
T.F.\ would like to thank the organizers of {\it QCD@Work~2007} for a lively and fruitful
workshop atmosphere in Martina Franca.
T.F.\ is supported by the German Ministry of Research
(BMBF, contract No.~05HT6PSA).
G.B.\ is supported by the DFG Sonderforschungsbereich/Transregio 9.
\end{theacknowledgments}



\begin{thebibliography}{99}


\bibitem{Gronberg:1997fj}
  J.~Gronberg {\it et al.}  [CLEO Collaboration],
  Phys.\ Rev.\  D {\bf 57}, 33 (1998).

\bibitem{Ball:1998je}
  P.~Ball,
  JHEP {\bf 9901}, 010 (1999).

\bibitem{Khodjamirian:2004ga}
  A.~Khodjamirian, T.~Mannel and M.~Melcher,
  Phys.\ Rev.\  D {\bf 70}, 094002 (2004).

\bibitem{Ball:2006wn}
  P.~Ball, V.~M.~Braun and A.~Lenz,
  JHEP {\bf 0605}, 004 (2006).

\bibitem{lattice}
  V.~M.~Braun {\it et al.},
  Phys.\ Rev.\  D {\bf 74}, 074501 (2006);
  A.~J\"uttner (UKQCD collab.), talk at {\it DA 06}, Durham (2006);
  L.~Del Debbio,
  Few Body Syst.\  {\bf 36}, 77 (2005).


\bibitem{Grozin:1996pq}
A.~G.~Grozin and M.~Neubert,
Phys.\ Rev.\ D {\bf 55}, 272 (1997).

\bibitem{Lange:2003ff}
B.~O.~Lange and M.~Neubert,
Phys.\ Rev.\ Lett.\  {\bf 91}, 102001 (2003).

\bibitem{Braun:2003wx}
V.~M.~Braun, D.~Y.~Ivanov and G.~P.~Korchemsky,
Phys.\ Rev.\ D {\bf 69}, 034014 (2004).



\bibitem{Ma:2006hc}
  J.~P.~Ma and Z.~G.~Si,
  Phys.\ Lett.\  B {\bf 647}, 419 (2007).


\bibitem{Braguta:2006wr}
  V.~V.~Braguta, A.~K.~Likhoded and A.~V.~Luchinsky,
  Phys.\ Lett.\  B {\bf 646}, 80 (2007);
  V.~V.~Braguta,
  Phys.\ Rev.\  D {\bf 75}, 094016 (2007);
  V.~V.~Braguta, Talk presented at this conference.



\bibitem{Bell:2006tz}
  G.~Bell, PhD thesis, LMU Munich 2006,
  arXiv:0705.3133 [hep-ph]; G.~Bell and T.~Feldmann,
  Nucl.\ Phys.\ Proc.\ Suppl.\  {\bf 164}, 189 (2007).

\bibitem{Bell:2007}
G.~Bell and T.~Feldmann,
Preprint SI-HEP-2007-20 (to be published).


\bibitem{Braun:1989iv}
  V.~M.~Braun and I.~E.~Filyanov,
  Sov.\ J.\ Nucl.\ Phys.\  {\bf 52}, 126 (1990).


\bibitem{Beneke:2000wa}
M.~Beneke and T.~Feldmann,
Nucl.\ Phys.\ B {\bf 592}, 3 (2001).



\bibitem{Kawamura:2001jm}
  H.~Kawamura, J.~Kodaira, C.~F.~Qiao and K.~Tanaka,
  Phys.\ Lett.\  B {\bf 523}, 111 (2001)
  [Erratum-ibid.\  B {\bf 536}, 344 (2002)].

\bibitem{ERBL}
A.~V.~Efremov and A.~V.~Radyushkin,
Phys.\ Lett.\ B {\bf 94}, 245 (1980);
G.~P.~Lepage and S.~J.~Brodsky,
Phys.\ Lett.\ B {\bf 87}, 359 (1979);
  G.~P.~Lepage and S.~J.~Brodsky,
  Phys.\ Rev.\  D {\bf 22}, 2157 (1980).





\bibitem{Braaten:1995ej}
  E.~Braaten and S.~Fleming,
  Phys.\ Rev.\  D {\bf 52}, 181 (1995).



\bibitem{Ball:2005ei}
P.~Ball and A.~N.~Talbot,
JHEP {\bf 0506}, 063 (2005).



\bibitem{Neubert:1993mb}
  M.~Neubert,
  Phys.\ Rept.\  {\bf 245}, 259 (1994).


\bibitem{Lee:2005gz}
  S.~J.~Lee and M.~Neubert,
  Phys.\ Rev.\ D {\bf 72}, 094028 (2005).



\bibitem{DeFazio:2005dx}
F.~De Fazio, T.~Feldmann and T.~Hurth,
Nucl.\ Phys.\  B {\bf 733}, 1 (2006);
%
arXiv:0711.3999 [hep-ph].







\end{thebibliography}
\end{document}